\documentclass[11pt,a4paper]{article}

\usepackage{graphicx}

\usepackage{authblk}

\usepackage{epstopdf}

\usepackage{cite}

\usepackage{hyperref}

\usepackage{geometry}
\usepackage{amsmath,amssymb}
\usepackage{subfigure}

\newcommand{\be}{\begin{equation}}
\newcommand{\ee}{\end{equation}}

\newcommand{\bea}{\begin{eqnarray}}
\makeatletter
\newcommand*{\rom}[1]{\expandafter\@slowromancap\romannumeral #1@}
\makeatother
\newcommand{\eea}{\end{eqnarray}}

\newcommand{\ba}[1]{\begin{array}{#1}}

\newcommand{\ea}{\end{array}}

\begin{document}

\begin{titlepage}

\vspace{0.5cm}

\begin{center}
\Large \bf{Metric for Rotating object in Infrared Corrected Nonlocal Gravity Model}
\end{center}

\vspace{0.3cm}

{ 
\begin{center}
\sc{Utkarsh Kumar, Sukanta Panda, Avani Patel}
\let\thefootnote\relax\footnotetext{ \hspace{-0.5cm} E-mail: {$\mathtt{utkarshk@iiserb.ac.in, sukanta@iiserb.ac.in, avani@iiserb.ac.in} $}}
\end{center}
}
\vskip 0.3 cm
\begin{center}
Indian Institute of Science Education and Research, Bhopal,\\ Bhauri, Bhopal 462066,\\
Madhya Pradesh, India\\
\end{center}





\begin{abstract}
Here, we derive the metric for the spacetime around rotating object for the gravity action having nonlocal correction of $R\Box^{-2} R $ to the Einstein-Hilbert action. Starting with the generic stationary, axisymmetric metric, we solve the equations of motion in linearized gravity limit for the modified action including energy-momentum tensor of the rotating mass. We also derive the rotating metric from the static metric using the Demanski-Janis-Newmann algorithm. Finally, we obtain the constraint on the value of $M$ by calculating the frame dragging effect in our theory and comparing it with that of General Relativity and Gravity Probe B results, where $M$ is the mass scale of the theory.  
\end{abstract}

\end{titlepage}

\section{Introduction}
The framework of the effective field theory of the Quantum Gravity predicts nonlocal correction terms in the gravity action. The effects of these nonlocal terms correspond to both IR and UV behaviour of gravity. The cosmological effect of nonlocal terms added into the Einstein-Hilbert(EH) action in the late time era are poorly studied. As it is well known that the cosmic microwave background and supernova data suggest that our universe has gone through two accelerating expansion phases in the early universe as well as in the recent era respectively known as Inflation and Late-time acceleration. Therefore, explanation of Inflation and Late-time acceleration are the primary motivations for the study of any nonlocal correction in the cosmological context. 



The correction term required to explain the accelerating expansion of our universe at recent times has to dominate at large distance. Nonlocal correction term in the EH action was first proposed by Wetterich\cite{Wetterich:1997bz}. However, this model does not produce the desired cosmological evolution. As the first attempt in this direction, in 2008, Deser and Woodard introduced a class of models considering a general nonlocal term of $Rf(\Box^{-1}R)$. The functional form of $f(\Box^{-1}R)$ is obtained by fitting it with the supernova data. Then structure formation for this particular functional form has been studied by several authors \cite{Woodard:2014iga,Barvinsky:2003kg,Barvinsky:2011hd,Barvinsky:2011rk,Dirian:2014xoa,Nersisyan:2016hjh,Amendola:2017qge,Foffa:2013vma,Kehagias:2014sda,Tsamis:2016boj,Tsamis:2005hd,Tsamis:2009ja,Deser:2013uya,Kumar:2018chy}.

In this work, we consider the RR model developed in \cite{Foffa:2013vma} which contains nonlocal term like $R\Box^{-2}R$. This nonlocal term mimics the cosmological constant at late times. Such nonlocal terms also appear in the effective action of pure gravity theory\cite{Codello:2016xhm,Codello:2016elq,Codello:2016neo}.

In 1918, Lense and Thirring derived a metric which describes the spacetime structure outside the rotating object which is known as Lense-Thirring metric\cite{LT:1918,Gravitation}. One can obtain the Lense-Thirring metric by taking weak field and slow rotation limit to the Kerr metric. The progress of the work presented here is similar to that of \cite{Cornell:2017irh} wherein the rotating metric (Lense-Thirring metric) is derived for the non-singular infinite derivative gravity. In this work, we derive the Lense-Thirring metric for RR model of nonlocal gravity taking two different approaches. We start with writing field equation for the model in linearized gravity limit which assumes the perturbative field $h_{\mu\nu}$ on top of the background Minkowski metric $\eta_{\mu\nu}$. This field equation is solved for different components of the general rotating metric. Thus achieved metric is characterized by two scalar potential (denoted by $\Phi$ and $\Psi$) and one vector potential. In General relativistic limit, two scalar potentials are same which is nothing but the Newtonian potential. We also derive the same metric using Demanski-Janis-Newmann (DJN) algorithm\cite{nj,Demianski:1972uza}.

The Kerr metric in General Relativity(GR) is singular at $r=0$ and $\theta=\pi/2$ where ($r,\theta,\phi$) are Boyer-Lindquist coordinates. This is called ring singularity since $r=0$ and $\theta=\pi/2$ correspond to equation of a ring $x^2+y^2=a^2$ and $z=0$ in Cartesian coordinates\cite{Hawking:1973uf}. We find that the rotating metric in nonlocal gravity model considered here also bears the ring singularity at $r=0$ and $\theta=\pi/2$.

Since GR has been proven to be correct in all experiments till date, any modification to GR needs to be checked against the experimental data. We calculate the geodetic precession and Lense-Thirring precession in an orbit around the Earth considering the rotating metric obtained here for RR model. We constrain the value of mass scale present as a coefficient of $R\Box^{-2}R$ term in the action considered by comparing the calculated values of geodetic and Lense-Thirring precession with Gravity Probe B satellite data~\cite{Everitt:2011hp}.

This paper is organized as follows: In Sec~\ref{sec:model}, we review the RR model of the nonlocal gravity and write the field equations for the model. In Sec.~\ref{sec:ll}, we derive the field equations in linearized gravity limit. The Lense-Thirring metric or rotating metric for the model is derived in Sec.~\ref{sec:rotatingmetric}. The analysis of the ring singularity for the metric obtained in Sec.~\ref{sec:rotatingmetric} is done in Sec.~\ref{sec:rs}. We calculate the geodetic precession and Lense-Thirring precession for the above mentioned metric and compare it with experimental values of both precessional motion observed by Gravity Probe B satellite in Sec.~\ref{sec:frmdrag}. Finally, we conclude our work in Sec.~\ref{sec:conclusion}

\section{ The Model and basic equations}\label{sec:model}
Let us begin with the  RR nonlocal model specified by following action
\begin{eqnarray}
S = \frac{1}{2\kappa^{2}}\int d^{4}x \sqrt{-g}\Big[ R + \frac{M^{2}}{3} R\frac{1}{\Box^2}R\Big] + \mathcal{L}_{m}.  \label{action}
\end{eqnarray}
Here M is the mass scale associated with nonlocal correction to the EH action. The nonlocality is denoted by the inverse of the d'Alembertian operator and $ \mathcal{L}_m $ is matter part of action. In the limit $M \rightarrow 0 $, the above action \eqref{action} reduces to Einstein- Hilbert action. The extensive study of the action under consideration has been done in \cite{Nersisyan:2016hjh,Dirian:2014xoa,Calcagni:2010ab}.

Equations of motion \cite{Edholm:2016hbt,Conroy:2014eja,Biswas:2013cha,Biswas:2013kla} corresponding to action~\eqref{action} are 

\begin{equation}
\begin{split}
2 \kappa^{2} T_{\alpha \beta} & = G_{\alpha \beta} + \frac{2}{3} M^2 G_{\alpha \beta} \frac{1}{\Box^{2}} R + \frac{2M^{2}}{3}g_{\alpha \beta} R \frac{1}{\Box^{2}} R - \frac{2}{3} M^2 ( \nabla_{\alpha} \nabla_{\beta} + g_{\alpha \beta} \Box)  \frac{1}{\Box^{2}} R \\& + \frac{2}{3} M^{2} \nabla_{(\alpha} R^{-1} \nabla_{\beta)} R^{-2} - \frac{2M^2}{3} g_{ \alpha \beta }( \nabla_{(\mu} R^{-1} \nabla^{\nu)} R^{-2} + 2 R^{-2}). \label{gumu}
\end{split}
\end{equation}
Another equivalent form of field equations can be obtained by the redefining the operation of inverse of d'Alembertian action on $R$ as scalar field and converting \eqref{action} into scalar-tensor action. That approach gives rise to following equation of motion
\begin{equation}
\begin{split}
\kappa^{2} T_{\alpha \beta} & =  G_{\alpha \beta} - \frac{M^{2}}{3}\Big\{ 2\Big(G_{\alpha
 \beta} - \nabla_{\alpha}\nabla_{\beta} + g_{\alpha \beta} \Box \Big)  + g_{\alpha \beta} \nabla^{\lambda}U \nabla_{\lambda}S  \\& - \nabla_{(\alpha}U \nabla_{\beta)}S  -\frac{1}{2}g_{\alpha \beta} U^{2} \Big\}
 \end{split}
\end{equation}
where $ U = - R\frac{1}{\Box}$  and $ S = - U\frac{1}{\Box}$.


\subsection{Linearized limit}\label{sec:ll}
 We consider the weak field limit of the field equation derived in Eq.~\eqref{gumu}. In weak field limit we take $g_{\mu\nu} $ as a perturbed metric around Minkowski background $ \eta_{\mu\nu} $ by a small amount,
\begin{equation}
g_{\mu\nu} = \eta_{\mu\nu} + h_{\mu\nu},\;\;\;|h|\ll 1 \label{pergmu}
\end{equation}
Using Eq.~\eqref{pergmu}, one can find the expressions for Riemann Tensor , Ricci Tensor and Ricci scalar as follows
\begin{eqnarray}
 R_{\rho\mu\sigma\nu} &=& \frac{1}{2} \Big( \partial_{\sigma}\partial_{\mu} h_{\rho\nu} + \partial_{\nu}\partial_{\rho} h_{\mu\sigma} -\partial_{\nu}\partial_{\mu} h_{\rho\sigma} -\partial_{\sigma}\partial_{\rho} h_{\sigma\mu}\Big) , \label{weakrt} \\
R_{\mu\nu}& =& \frac{1}{2}\Big( \partial^{\sigma}\partial_{\mu}h_{\sigma \nu} + \partial_{\nu}\partial_{\sigma} h_{\mu}^{\sigma} -\partial_{\nu}\partial_{\mu}h -\Box h_{\mu\nu} \Big) ,  \label{weakri} \\ 
R &= &\partial_{\mu}\partial_{\nu}h^{\mu\nu} - \Box h . \label{weakrs}
\end{eqnarray}
Then the field equation \eqref{gumu} becomes
\begin{equation}
\begin{split}
2\kappa^2 T_{\mu\nu} &= -\Bigg[\Box h_{\mu\nu} -\partial_{\sigma} \partial_{(\mu}h^{\sigma}_{\nu)} + \Big(1  - \frac{2M^{2}}{3} \Box^{-1}  \Big)(\partial_{\mu}\partial_{\nu}h + \eta_{\mu\nu}\partial_{\alpha}\partial_{\beta}h^{\alpha\beta}) \\&-\Big( -1  + \frac{2M^2}{3} \Box^{-1}  \Big) \eta_{\mu\nu} \Box h  + \frac{2M^{2}}{3} \Box^{-2}\nabla_{\mu}\nabla_{\nu}\partial_{\alpha}\partial_{\beta}h^{\alpha\beta}\Bigg].
\end{split}    \label{rearrgmunu}
\end{equation}


\section{Metric for Rotating object}\label{sec:rotatingmetric}
In this section, we calculate the spacetime metric in the exterior region of the rotating object for the model considered in Eq.~\eqref{action}. Consider the generic rotating metric as
\begin{equation}
ds^{2} = -(1 + 2\Phi ) dt^2 + 2 \vec{h}.d \mathtt{x} dt + (1 - 2\Psi)d \mathtt{x}^2.
\label{metric1}
\end{equation}
In the case of GR, $\Phi=\Psi$ is the Newtonian potential. Note that the components of $h_{\mu\nu}$ are
\begin{eqnarray}
h_{00} &=& -2\Phi  \\
h_{ij} &=&  -2\Psi \eta_{ij}  \\  \vec{h} &=& h_{0x} \hat{x} + h_{0y} \hat{y} + h_{0z} \hat{z}. 
\end{eqnarray}  
The components of stress-energy tensor for the rotating object having energy density $\rho=m\delta^3(\vec{r})$ with mass $m$ and angular velocity $v_i$  are given by
\begin{eqnarray}
T_{00} &=& \rho \\ T_{0i} & = & -\rho v_{i}.\label{sttensor} 
\end{eqnarray}
Notice that the rotation of the object explains the presence of the angular momentum terms $T_{0i}$ in the stress-energy tensor. Taking the trace of linearized field equation \eqref{rearrgmunu} and substituting the metric \eqref{metric1} and stress-energy tensor components given in \eqref{sttensor}, we obtain
\begin{equation}
\begin{split}
\rho = -2(1 - M^{2}\Box^{-1})(\Box h -\partial_{\alpha}\partial_{\beta}h^{\alpha \beta}). \label{traceeqn}
\end{split}
\end{equation}
Using equations (\ref{rearrgmunu}) and (\ref{traceeqn}) we obtain the equations of motion for $h_{00}$ , $h_{ij}$ and $h_{0i}$ as
\begin{eqnarray}
\rho & = & 4(1 - M^{2}\Box^{-1})(\nabla^{2}\Phi -2 \nabla^{2}\Psi), \label{oocom} \\
\rho &=& -\frac{4}{3}M^{2}\Box^{-1}(\nabla^{2}\Phi -2 \nabla^{2}\Psi) - 4\nabla^{2} \Psi, \label{11com} \\
\kappa \rho v_i &=&  -2\nabla^{2}h_{0i}.  \label{oi comp}
\end{eqnarray}
One can see that there is no off-diagonal terms $h_{0i}$ in the Eqns.~\eqref{oocom} and \eqref{11com}. It is also apparent that the Eq.~\eqref{oi comp} is unaffected by the non-local gravity correction and has the same form as in GR. Solving equations (\ref{oocom}) - (\ref{oi comp}) we get
\begin{eqnarray}
\Phi(r) &=& \frac{m}{24\pi M_{p}^{2}r}(e^{-Mr}-4) \longrightarrow \frac{Gm}{r}\Big(\frac{e^{-Mr}-4}{3}\Big), \label{phipot} \\
\Psi(r) & = & \frac{m}{24\pi M_{p}^{2}r}(-e^{-Mr}-2) \longrightarrow \frac{Gm}{r}\Big(\frac{-e^{-Mr} - 2}{3}\Big), \label{psipot} \\
h_{0x} & = & -\frac{m v_{x}}{2\pi M_{p}^{2}r} \longrightarrow -\frac{4Gm v_{x}}{r}, \label{hoxpot} \\
h_{0y} & = & -\frac{m v_{y}}{2\pi M_{p}^{2}r} \longrightarrow -\frac{4Gm v_{y}}{r}, \label{hoypot} \\
h_{0z} & = & -\frac{m v_{z}}{2\pi M_{p}^{2}r} \longrightarrow  -\frac{4Gm v_{z}}{r}. \label{hozpot} 
\end{eqnarray}
We assume the case where source is moving in such direction so that its angular momentum points in the $z$ direction. Therefore, we can write the velocities as follows 
\begin{equation}
\begin{split}
v_{x} = -y \omega \hspace{1em},\hspace{1em} v_{y} = x \omega \hspace{1em},\hspace{1em} v_{z} = 0 \hspace{2em}.
\end{split}
\end{equation}
From equations \eqref{phipot} - \eqref{psipot}, we rewrite equations \eqref{hoxpot} - \eqref{hozpot} as
\begin{eqnarray}
h_{0x} = -2y \omega (\Phi(r) + \Psi(r)) \hspace{1em},\hspace{1em} h_{0y} = 2x \omega (\Phi(r) + \Psi(r)),
\end{eqnarray}
The resulting metric is given by
\begin{equation}
ds^{2} = -(1 + 2\Phi)dt^{2} + 4(\Phi + \Psi)(x\omega dt dy - y \omega dt dx) + (1 - 2\Psi) d\mathtt{x}^2.
\end{equation}
Furthermore, we can convert the above metric from Cartesian coordinates to Boyer-Lindquist coordinates (t, r, $\theta$, $\phi $) via the transformations
\begin{equation}
\begin{array}{rl}
x=&\sqrt{r^2+a^2}\,{\rm sin}\theta\,{\rm cos}\phi,\\
y=&\sqrt{r^2+a^2}\,{\rm sin}\theta\,{\rm sin}\phi,\\
z=&r{\rm cos}\theta.
\end{array}
\end{equation}
and as a result we get 
\begin{equation}
\begin{split}
ds^{2}   = & -(1 + 2\Phi)dt^{2} +4\frac{J\sin^{2}\theta}{m}(\Phi + \Psi) d\phi dt +\\& (1- 2\Psi)( dr^2 + r^{2}d\theta^{2} + r^{2}\sin^{2}\theta d\phi^{2}),  \label{rotmet}
\end{split}
\end{equation}

where J is angular momentum, defined as $ v = \frac{r \times J }{m r^{2}}$. This is a case of very slowly rotating object and therefore we take $r^2+a^2\sim r^2$ in the transformation equations. If we take $r\rightarrow\infty$ limit then the metric in Eq.~\eqref{rotmet} reduces to the rotating metric in GR which shows that the non-local gravity correction attenuates at very large distances. 


\subsection{Rotating metric from DJN Algorithm}
The Demanski-janis-Newmann algorithm provides a scheme to transform static metric into rotating metric using complex coordinate transformations.\cite{nj,Demianski:1972uza,Erbin:2014aja,Erbin:2014aya}. In this section, we apply this algorithm on the weak gravity static metric to reproduce the metric in Eq.~\eqref{rotmet}. First, we write the static metric
\begin{equation}
ds^{2} = -(1 + 2\Phi ) dt^2  + (1 - 2\Psi)d \mathtt{x}^2.  \label{stat}
\end{equation}
Transforming the above metric into spherical polar coordinates, one can rewrite metric \eqref{stat} in the following form
\begin{equation}
ds^{2} = -f_{t}dt^{2} + f_{r}dr^{2} + f_{\Omega}(d \theta^2 + \sin^{2}\theta d\phi^{2} ),   \label{spec}
\end{equation}
where $ f_t =  1+ 2\Phi$ , $ f_{r} = 1-2\Psi $ and $ f_{\Omega} = r^{2}f_r $.

Performing the null coordinate transformation~\cite{nj,Demianski:1972uza,Erbin:2014aja,Erbin:2014aya} $ t =  u + (\frac{1-2\Psi}{1-2\Phi})^{1/2}r$, Eq.~\eqref{spec} yields
\begin{equation}
ds^2 = -(1-2\Phi) du^2 -2\sqrt{(1-2\Phi)(1-2\Psi)}dudr + f_{\Omega}d\Omega^{2}  \label{null}
\end{equation}
where $ d \Omega^{2} = d \theta^{2} + \sin^{2}\theta d\phi^{2}  $.
The next step is to complexify the coordinates in metric \eqref{null} as following 
\begin{eqnarray}
r\longrightarrow r' = r + a\mathit{i}\cos \theta ,  \hspace{1em} u\longrightarrow u' = u - a\mathit{i}\cos \theta.   \label{complex}
\end{eqnarray}
In Eq.~\eqref{complex}, we introduced rotation parameter $a$ defined by $ a \equiv \frac{J}{m} $. 
Using the above transformations and the ansatz $ i d\theta = \sin \theta d \phi $, we obtain the differential transformations of r and u as
\begin{eqnarray}
dr &=& dr' - a\sin ^{2}\theta d\phi \\  du &=& du' + a\sin ^{2}\theta d\phi.
\end{eqnarray}
In the DJN approach we have to be careful while choosing transformations for r, $ r^{2} $ and $ \frac{1}{r} $ that the functions $f_{i}$'s remain real and their angle dependence should be of $ \cos \theta $. Ensuring this, the transformations we find are 
\begin{eqnarray}
r & \longrightarrow & r' \\
\frac{1}{r} & \longrightarrow & \frac{Re(r')}{|r'|^{2}} \\
r^2 &\longleftrightarrow & |r'|^{2}.
\end{eqnarray}
Therefore our functions become
\begin{eqnarray}
f_{t}(r) \longrightarrow \tilde{f_{t}}(r,\theta) = 1 + \frac{mr}{24 \pi M_{p}^{2}\Sigma} \Big( e^{-mr} - 4 \Big)   \label{phitilde}   \\
f_{r}(r) \longrightarrow \tilde{f_{r}}(r,\theta) = 1 + \frac{mr}{24 \pi M_{p}^{2}\Sigma} \Big( e^{-mr} + 2 \Big),  
 \label{psitilde} \\
 r^{2} \longrightarrow  \Sigma \equiv r^{2} + a^{2}  \cos^{2}\theta \hspace{8em}& &.
\end{eqnarray}
then writting down the null rotating metric
\begin{equation}
ds^{2} = -\tilde{f_{t}}( du + \alpha dr + \omega \sin \theta d\phi )^{2} + 2\beta dr d\phi + \Sigma \tilde{f_{r}}(d\theta^{2} + \sigma^{2} \sin^{2}\theta d\phi^{2}), 
\end{equation}
where, 
\begin{eqnarray}
\omega &=& a \sin \theta - \sqrt{\frac{\tilde{f}_r}{\tilde{f}_t}} a \sin \theta ,\\
\sigma^{2} & = & 1 + \frac{a^2 \sin^2 \theta}{r^{2} + a^{2}} ,\\
\alpha &=& \sqrt{\frac{\tilde{f}_r}{\tilde{f}_t}} , \\
\beta & =& -\tilde{f}_r a \sin^{2} \theta .
\end{eqnarray}
The last step is to convert null metric into Boyer-Lindquist form. To do that we have to make sure that 
\begin{eqnarray}
g(r) & =& \frac{\sqrt{(\tilde{f}_t\tilde{f}_r)^{-1}}\tilde{f}_{\Omega} - F'G'}{\Delta}, \\ h(r) &=& \frac{F'}{H(\theta) \Delta}
\end{eqnarray}
are functions of $r$ only where $\Delta=(\tilde{f_{\Omega}}/\tilde{f_r})\sigma^2$. It is true provided $ \Phi \ll 1 ,  $ such that $ f_{r}^{-1} = f_{t} $. These transformations are valid only when we consider very small perturbation around Minkowski background. After some trivial algebra we obtain 
\begin{equation}
\begin{split}
ds^{2} = & -(1 + 2\tilde{\Phi}) dt^{2} + 4a(\tilde{\Phi} + \tilde{\Psi}) \sin^{2}\theta d\phi dt + \frac{\Sigma(1 - 2\tilde{\Psi})}{r^2 + a^2}dr^2 \\& +\Sigma (1 - 2\tilde{\Psi})\Big( d\theta^{2} + sin^{2}\theta \Big( \frac{r^{2} + a^{2}}{\Sigma}\Big) d\phi^{2}\Big),
\label{metric3}
\end{split}
\end{equation}
where 
\begin{eqnarray}
\tilde{\Phi} &=&  \frac{mr}{24 \pi M_{p}^{2}\Sigma} \Big( e^{-Mr} - 4 \Big)  \\
 \tilde{\Psi}&=&\frac{mr}{24 \pi M_{p}^{2}\Sigma} \Big(- e^{-Mr} - 2 \Big).
\end{eqnarray}


\section{Ring Singularity}\label{sec:rs}

%
%
%
%
%
%

To find the ring singularity in the metric \eqref{metric3} we follow the same procedure as in \cite{Buoninfante:2018xif}. Let us consider a rotating ring having mass $m$ and radius $a$. The ring lies in the X-Y plane with $z=0$ and the angular velocity of the ring points in the direction of Z-axis. The $(00)$-component of the energy momentum tensor of the source is given by
\begin{equation}
T_{00}=m\delta(z)\frac{\delta(x^2+y^2-a^2)}{\pi}. \label{ring}
\end{equation}
The above distribution for the energy-momentum tensor is similar to that of the distributional form of the energy-momentum tensor of the Kerr metric \cite{Balasin:1993kf}. We also have the following non-vanishing components of the stress-energy tensor:
\begin{equation}
T_{0i}=T_{00} v_i,
\end{equation}
where $v_i$ is the same angular velocity as defined earlier.
%
%

Now we rewrite the general linearized metric given in Eq.~\eqref{metric1}

\begin{equation}
ds^2=-(1+2\Phi)dt^2+2 \vec{h}\cdot d\vec{x}dt+(1-2\Psi)d\vec{x}^2,\label{metric}
\end{equation}
%
Differential equations for the each component of metric are given by
\begin{equation}
\begin{array}{rl}
\displaystyle \frac{(\Box + M^2)}{(3\Box + 4M^2)}\,\nabla^2 \Phi(\vec{r})  = & 4Gm\, \delta(z)\,\delta(x^2+y^2-a^2),\\
\displaystyle \frac{(\Box + M^2)}{(3\Box + 2M^2)}\,\nabla^2 \Psi(\vec{r})  = & 4Gm\, \delta(z)\,\delta(x^2+y^2-a^2),\\
\displaystyle \nabla^2 h_{0x}(\vec{r})=& \displaystyle-8Gm\omega y\, \delta(z)\,\delta(x^2+y^2-a^2),\\
\displaystyle \nabla^2 h_{0y}(\vec{r})=& \displaystyle 8Gm\omega x\, \delta(z)\,\delta(x^2+y^2-a^2).
\end{array}\label{diff-eq}
\end{equation}
In the next section, we solve equations \eqref{diff-eq} and examine the presence of ring singularity. Before going to the next section, let us look at the Kerr metric given by \cite{Visser:2007fj}
\begin{equation}
ds^{2}=  -\left(1-\dfrac{2mr}{\Sigma}\right)dt^{2}-\dfrac{4mar{\rm sin}^{2}\theta}{\Sigma}dtd\phi+\dfrac{\Sigma}{\Delta}dr^{2}+\Sigma d\theta^{2} + {\rm sin}^{2}\theta\left(r^{2}+a^{2}+\dfrac{2ma^{2}r{\rm sin^{2}}\theta}{\Sigma}\right)d\phi^{2},
\label{metric-1}
\end{equation}
where, $\Sigma\equiv r^2+a^2\rm{cos}^2\theta$ and $\Delta\equiv r^2-2mr+a^2$ with $a$ being the rotation parameter. It is easy to observe that the Kerr metric in \eqref{metric-1} becomes singular (see Ref. \cite{Hawking:1973uf}) when $\Sigma$ becomes zero. The Kretschmann scalar blows up as $\Sigma=0$ at $r=0$ and $\theta=\pi/2,$ which in Cartesian coordinates means \cite{Hawking:1973uf} 
\begin{equation}
x^2+y^2=a^2,\,\,\,\,\,\,z=0,\label{cartesian-ring}
\end{equation}
which is nothing but the equation of a ring of radius $a$. Thus GR admits the ring singularity in the Kerr spacetime.

\subsection{Computation of metric component $ h_{00}$, $ h_{0i}$, $ h_{ij}$}
Solving differential equations written in Eqns.\eqref{diff-eq} and obtaining expressions for the metric components is easier in momentum space. First we find the fourier transform for $ \delta(z)\delta(x^2+y^2-a^2) $ 
\begin{equation}
\mathcal{F}[\delta(z)\delta(x^2+y^2-a^2)]=\displaystyle \int dx
\,dy\,dz\, \delta(z)\,\delta(x^2+y^2-a^2)\, e^{ik_xx}e^{ik_yy}e^{ik_zz}.
\end{equation}
The above integral can be computed in cylindrical coordinates:
\begin{equation} 
x=\rho {\cos}\varphi,\,\,\,\,\,y=\rho {\rm sin}\varphi,\,\,\,\,\,z=z.
\end{equation}
We have,
\begin{equation}
\begin{array}{rl}
\mathcal{F}[\delta(z)\delta(x^2+y^2-a^2)]= & \displaystyle\int\limits_{-\infty}^{\infty} dz \delta(z) e^{ik_z z}\int\limits_{0}^{\infty}d\rho \rho \delta(\rho^2-a^2)  \int\limits_{0}^{2\pi}d\varphi  e^{ik_x\rho {\cos}\varphi}e^{ik_y\rho {\rm sin}\varphi}\\
=& \displaystyle \pi I_0\left(ia\sqrt{k_x^2+k_y^2}\right),\label{fourier00}
\end{array}
\end{equation}
where $I_0$ is a Modified Bessel function, which is also defined in terms of the Bessel function as $I_0(x)=J_0(ix)$.

Using result \eqref{fourier00} in differential equation for $ \Phi$  and taking the inverse fourier transform we get the following expression for newtonian potential $ \Phi  $
 \begin{equation}
 \Phi(\vec{r})= -4\pi Gm \int \frac{d^3k}{(2\pi)^3}\frac{(3k^2 + 4M^2)}{k^{2}(k^2 + M^2)} I_0\left(ia\sqrt{k_x^2+k_y^2}\right)e^{ik_xx}e^{ik_yy}e^{ik_zz}, \label{grav-pot}
 \end{equation}
where, $d^{3}k $ is differntial volume in momentum space. To study the ring singularity in our model of intrest, we restrict ourselves in the ring plane (i.e. x-y plane , z $ = 0$)  and transform the Eq.~\eqref{grav-pot} in cylindrical coordinates via following transformations
\begin{equation}
 \rm k_{x} = \zeta \rm cos \phi,\;\;\; k_{y} = \zeta \rm sin \phi,\;\;\; k_{z} = k_{z}, \label{cylcoordtransf} 
\end{equation}

then we have the final expression for $ \Phi( \rho)$ as 
\begin{equation}
\Phi(\rho)= -Gm \int\limits_{0}^{\infty}d\zeta I_0\left(ia\zeta\right)I_0\left(i\zeta \rho \right)\frac{(3k^2 + 4M^2)}{(k^2 + M^2)},
\label{grav-potIDGnum}
\end{equation}
which in the limit $M\rightarrow  0 $ gives the metric potential in the case of GR:
\begin{equation}
\Phi_{GR}(\rho)= - 3 Gm \int\limits_{0}^{\infty}d\zeta I_0\left(ia\zeta\right)I_0\left(i\zeta\rho\right).
\label{grav-potGRnum}
\end{equation}
similar expression can be found for $ \Psi$ as 
\begin{equation}
\Psi(\rho)= -Gm \int\limits_{0}^{\infty}d\zeta I_0\left(ia\zeta\right)I_0\left(i\zeta \rho \right)\frac{(3k^2 + 2M^2)}{(k^2 + M^2)}. \label{gr-psi}
\end{equation}

\begin{figure}[!ht]
  \centering
 \includegraphics[scale=0.4]{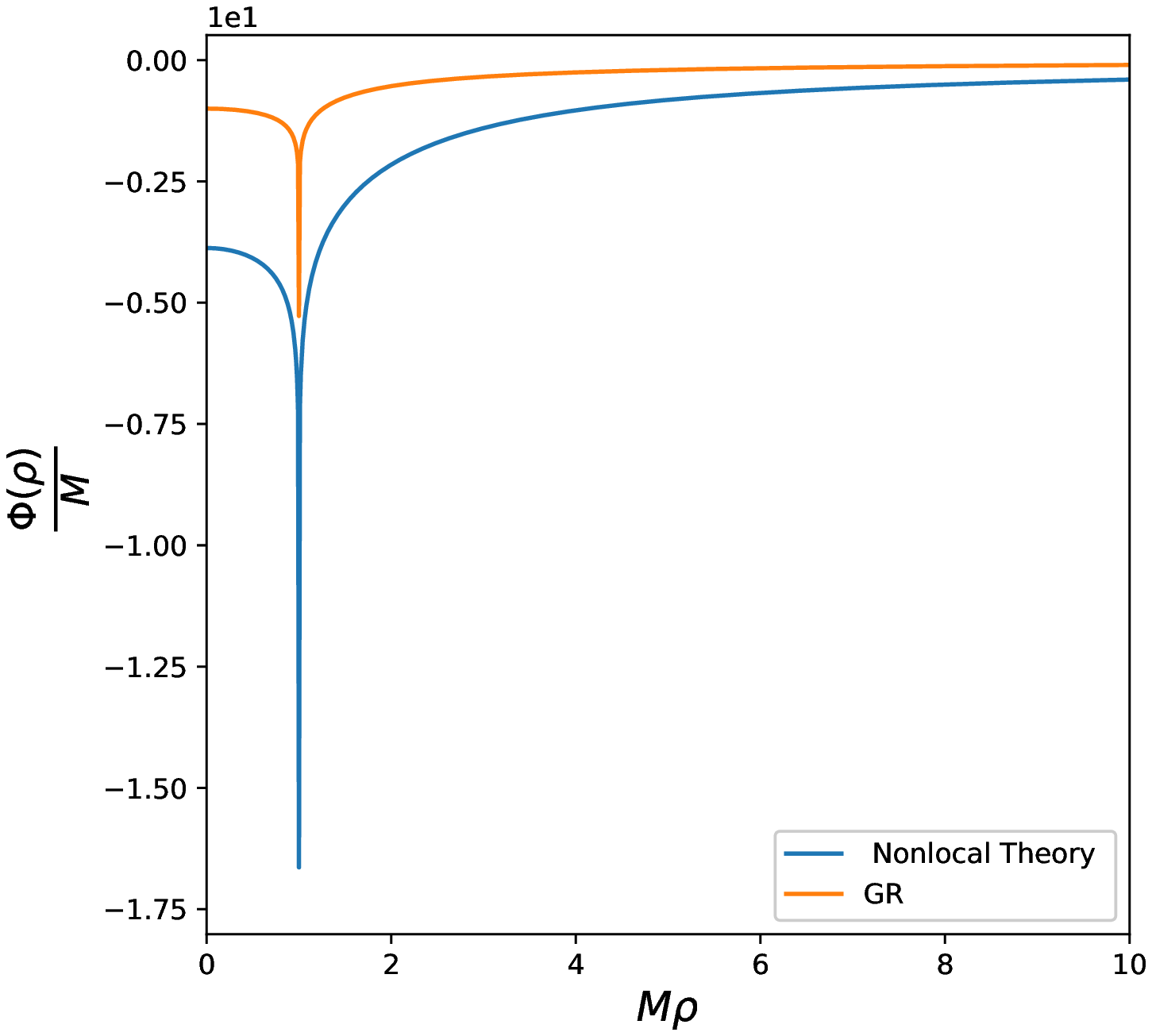} 
 \includegraphics[scale=0.4]{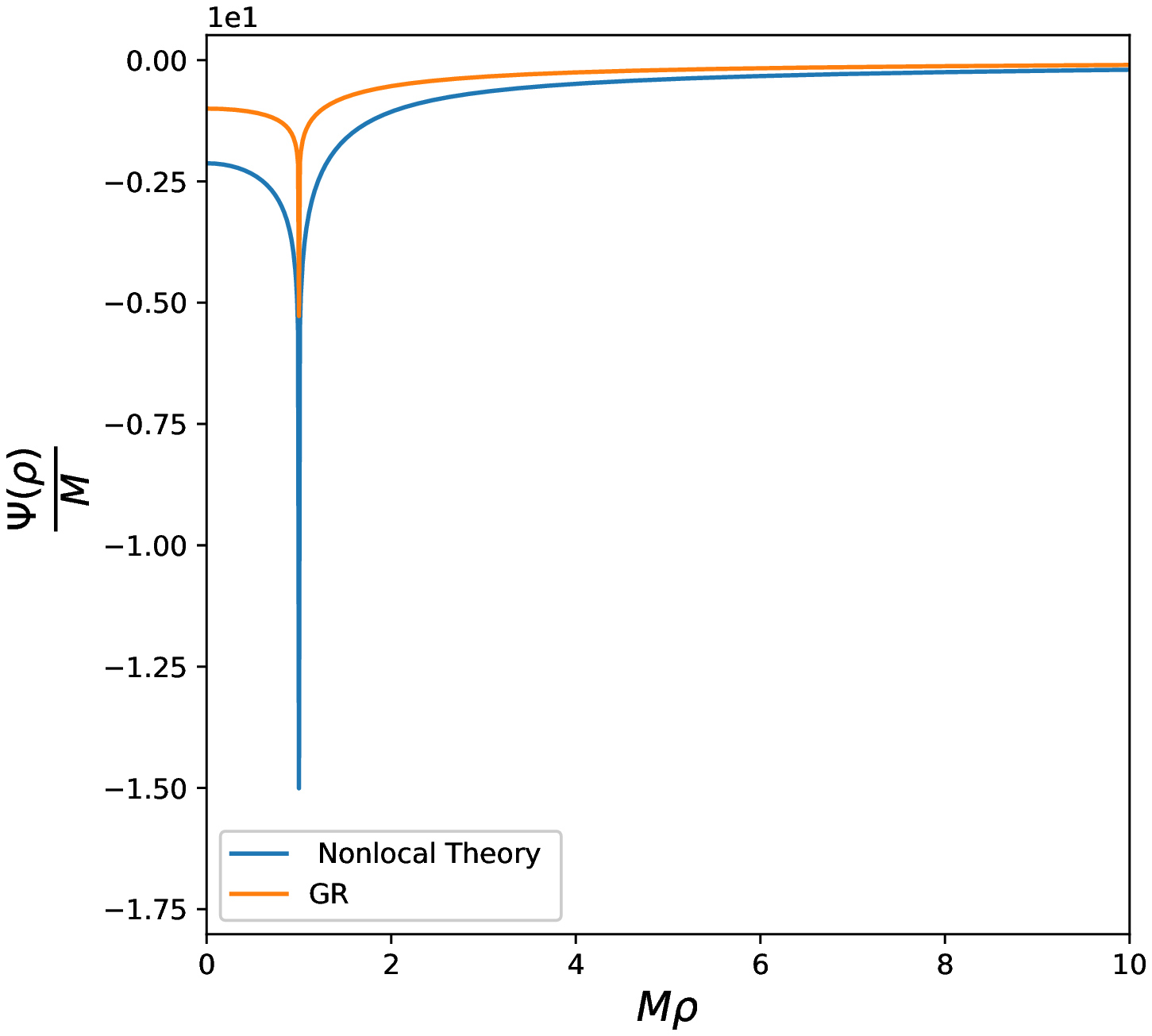} 
 \includegraphics[scale=0.4]{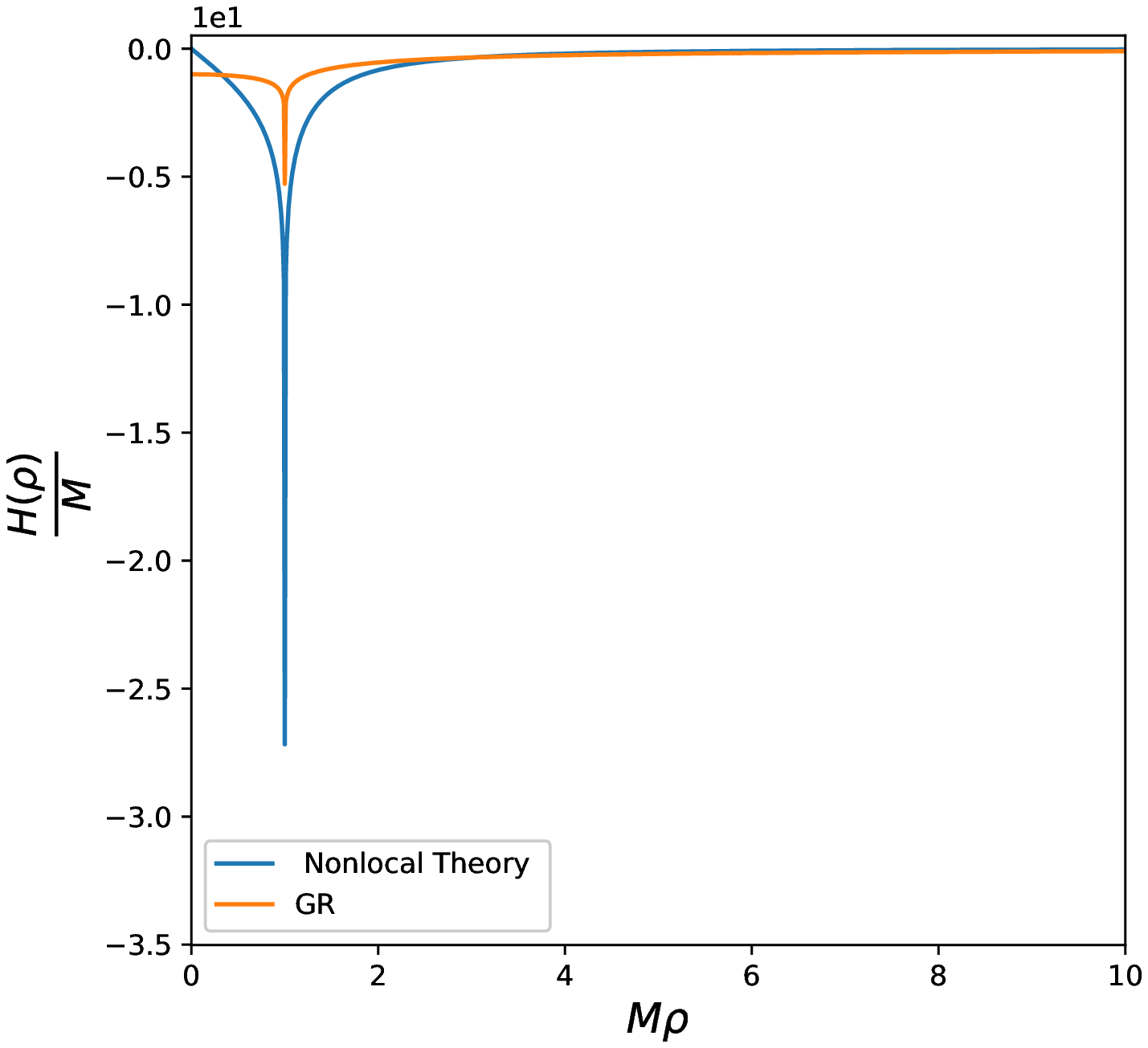}
\caption{Plots for $ \Phi(\rho)/M$ ,$ \Psi(\rho)/M$ and $H (\rho)/M $. We have chosen $Ma=1.0$.} \label{pab} 
\end{figure}

To compute the $ h_{0i}$, first we find the fourier transform of $ x \delta(z)\delta(x^2+y^2-a^2) $
\begin{equation}
\begin{array}{rl}
\mathcal{F}[x\delta(z)\delta(x^2+y^2-a^2)]= & \displaystyle\int\limits_{-\infty}^{\infty} dz \delta(z) e^{ik_z z}\int\limits_{0}^{\infty}d\rho \rho^2  \delta(\rho^2-a^2)  \int\limits_{0}^{2\pi}d\varphi  e^{ik_x\rho {\cos}\varphi}e^{ik_y\rho {\rm sin}\varphi}{\rm cos}\varphi\\
=& \displaystyle \pi  a \frac{k_x}{\sqrt{k_x^2+k_y^2}} I_1\left(ia\sqrt{k_x^2+k_y^2}\right),
\end{array}
\end{equation}
and similarly we also obtain
\begin{equation}
\mathcal{F}[y\delta(z)\delta(x^2+y^2-a^2)]=\pi a \frac{k_y}{\sqrt{k_x^2+k_y^2}} I_1\left(ia\sqrt{k_x^2+k_y^2}\right),
\end{equation}
then $h_{0i}$ are given by following expressions 
\begin{equation}
h_{0x}(\vec{r})= 16Gm\omega a \int \frac{d^3k}{(2\pi)^3}\frac{(6k^2 + 6M^2)}{k^2(k^2 + M^2)} \frac{k_x}{\sqrt{k_x^2+k_y^2}}I_1\left(ia\sqrt{k_x^2+k_y^2}\right)e^{ik_xx}e^{ik_yy}e^{ik_zz}, \label{int-h0x}
\end{equation}
\begin{equation}
h_{0y}(\vec{r})= -16Gm\omega a \int \frac{d^3k}{(2\pi)^3}\frac{(6k^2 + 6M^2)}{k^2(k^2 + M^2)} \frac{k_y}{\sqrt{k_x^2+k_y^2}}I_1\left(ia\sqrt{k_x^2+k_y^2}\right)e^{ik_xx}e^{ik_yy}e^{ik_zz}. \label{int-h0y}
\end{equation}
By using cylindrical coordinates and setting $z=0,$ we can obtain similar expressions for the cross-terms:
\begin{equation}
h_{0x}(x,y)= 2Gm\omega a \frac{y}{\rho}\int\limits_{0}^{\infty} d\zeta I_1(ia\zeta)I_1(i\zeta \rho) \frac{(6k^2 + 6M^2)}{(k^2 + M^2)},\label{h0x}
\end{equation}
\begin{equation}
h_{0y}(x,y)= -2Gm\omega a \frac{x}{\rho}\int\limits_{0}^{\infty} d\zeta I_1(ia\zeta)I_1(i\zeta \rho) \frac{(6k^2 + 6M^2)}{(k^2 + M^2)},\label{h0y}
\end{equation}
where remember that $\rho=\sqrt{x^2+y^2}$ is the radial cylindrical coordinate in the plane $z=0.$ Note that since $\theta=\pi/2$, we have
\begin{equation}
\frac{x}{\rho}={\rm cos}\phi,\,\,\,\,\,\,\frac{y}{\rho}={\rm sin}\phi,
\end{equation}
thus all the radial dependence and the singularity structure are taken into account by the following integral:
\begin{equation}
H(\rho):=\int\limits_{0}^{\infty} d\zeta I_1(ia\zeta)I_1(i\zeta \rho) \frac{(6k^2 + 6M^2)}{(k^2 + M^2)},\label{H-NLG}
\end{equation}
which in the limit $M\rightarrow 0$ gives the GR case:
\begin{equation}
H_{GR}(\rho):=\int\limits_{0}^{\infty} d\zeta I_1(ia\zeta)I_1(i\zeta \rho).\label{H-GR}
\end{equation}
Computation of integrals involved in expressions of graviationl potential $\Phi$ and $\Psi$ in Eqns.~\eqref{grav-potGRnum} and \eqref{gr-psi} and of $H(\rho)$ in Eq.~\eqref{H-NLG} is not possible analytically. Here, we solve them numerically and show the results in Fig.~\ref{pab}.
It is evident from the figure that the gravitational potential $\Phi$ and $\Psi$ and the off-diagonal component of the metric $h_{0i}$ show the singular characteristic at $M\rho=1$.

\section{Frame Dragging Effect}\label{sec:frmdrag}
The rotating metric is valid in weak field limit for any rotating astrophysical object which possesses the axial symmetry. Thus the metric derived in Eq.~\eqref{metric3} can be applied to the spacetime around the Earth, under the assumption that the underlying theory of gravity has nonlocal correction like given in Eq.~\eqref{action} to GR action. Any object which revolves around the Earth will experience two types of precessional motion of general relativistic origin. One is due to the geodetic drift and another is due to the frame dragging drift. The gravity probe B satellite has measured these two precessions, and they were compared to the predicted value from General Relativity. 

The geodetic precession is caused by the curvature of the spacetime due to the rotating object's mass and Lense-Thirring precession or frame dragging effect is the result of drifting of the frame due to the rotation of the object. As we can see in Eq.~\eqref{metric1}, the metric is composed of a scalar field $\Phi$ and a vector field $\vec{h}$. These scalar fields and vector field depend on the shape of the body and mass and velocity distribution of the body. These fields can be expanded in terms of multipole moments. For our calculation, we consider only monopole moment. The formulae for instantaneous geodetic precession and instantaneous Lense-Thirring  precession in Cartesian coordinates are given by\cite{Adler:1999yt}
\begin{equation}
\Omega_G=\frac{3}{2}\nabla\Phi \times \vec{V},\;\;\;\Omega_{LT}=\frac{1}{2}\nabla \times \vec{h},
\end{equation}
where $\vec{V}$ is the four velocity of the orbiting gyroscope and $h_{0i}$ are the off-diagonal terms of the rotating metric. The instantaneous geodetic precession and Lense-Thirring precession for nonlocal gravity theory can be expressed as
\begin{equation}
\Omega_{G(NL)}=\frac{1}{3}\, e^{-Mr}\,(-1+4e^{Mr}-Mr)\,\Omega_{G(GR)},       
\label{GTprecession_comp}
\end{equation}
\begin{equation}
\Omega_{LT(NL)}=\Omega_{LT(GR)}  ,
\label{LTprecession_comp}
\end{equation}
where general relativistic expressions of geodetic precession and Lense-Thirring precessions are given by
\begin{equation}
\Omega_{G(GR)}= \frac{3Gm}{2r^3}(\vec{r}\times\vec{V}),\;\;\;\Omega_{LT(GR)}=\frac{2G}{r^3}\vec{J}
\end{equation}
The Gravity Probe B satellite containing four gyroscopes was launched in 2004. These gyroscopes measured geodetic and frame-dragging precessions in orbit with altitude $r=650$ km from the surface of the Earth. The result of the measurements done by Gravity Probe B for the geodetic drift rate was $\Omega_G=6601.8\pm 18.3$ milliarcsec/year and for the frame-dragging drift rate was $\Omega_{LT}=37.2\pm 7.2$ milliarcsec/year\cite{Everitt:2011hp}. The GR predicted geodetic drift rate is $\Omega_{G(GR)}=6606.1$ milliarcsec/year and frame-dragging drift rate is $\Omega_{LT(GR)}=39.2$ milliarcsec/year.

One can constrain the value of $M$ by checking for what values of $M$, $\Omega_{G(NL)}$ and $\Omega_{LT(NL)}$ match with $\Omega_{G(GR)}$ and $\Omega_{LT(GR)}$ having the difference well within the error bars of Gravity Probe B results. Since $M$ comes multiplied with $R$ in Eq.~\eqref{GTprecession_comp}, we find constraint on $Mr$ which comes out to be $Mr\leq 0.117$. Considering the value of radial distance of Gravity Probe B satellite from the centre of the Earth which is $r=7021$ km, we can obtain converting the above constraint into the constraint on $M$ which is $M\leq 3.299\times 10^{-15}$ eV. This is reverse of the condition obtained in \cite{Cornell:2017irh} which can be justified in a way that IR corrections to GR are taken here while in \cite{Cornell:2017irh} UV corrections were considered. 


\section{Conclusions}\label{sec:conclusion}
In this work, we have derived the metric for the exterior spacetime of the rotating body starting from the general rotating metric in the modified gravity theory having nonlocal gravity corrections to the Einstein-Hilbert action. The rotating metric which we found in Eq.~\eqref{rotmet} reduces to GR form in large $r$ limit. We also found that the off-diagonal terms of the metric are unchanged from the rotating metric in GR. 

In the last section, we calculated the instantaneous geodetic precession and instantaneous Lense-Thirring precession of the satellite orbiting the Earth for the model considered in \eqref{action} using the rotating metric derived in Eq.~\eqref{rotmet}. We found that the instantaneous geodetic precession $\Omega_{G(NL)}$ for the model \eqref{action} differs from that of GR $\Omega_{G(GR)}$ by a multiplicative factor while instantaneous Lense-Thirring precession is same as in GR. We have compared the values of geodetic precession and Lense-Thirring precession with the Gravity Probe B satellite's data and put the constraint on the value of the scale $M$ which comes out to be $M\leq 3.299 \times 10^{-15}$ eV. 

The rotating metric obtained in this paper can be utilized further in the studies of tests of modified gravity theories using gravitational wave astronomy. In particular, one can consider the metric \eqref{rotmet} as Kerr metric with small perturbation and can calculate the deviations in the frequencies of the gravitational waves emitted by the test particle orbiting a super massive blackhole which consequently can be useful in modeling Extreme Mass Ratio Inspirals(EMRI) for the modified gravity with nonlocal gravity corrections. 


\section{Acknowledgement} This work was partially supported by DST grant no. SERB/PHY/2017041.



\begin{thebibliography}{99}
\bibitem{Wetterich:1997bz} 
  C.~Wetterich,
  ``Effective nonlocal Euclidean gravity,''
  Gen.\ Rel.\ Grav.\  {\bf 30}, 159 (1998)
  [gr-qc/9704052].

\bibitem{Woodard:2014iga} 
  R.~P.~Woodard,
  ``Nonlocal Models of Cosmic Acceleration,''
  Found.\ Phys.\  {\bf 44}, 213 (2014)
  [arXiv:1401.0254 [astro-ph.CO]].
  
\bibitem{Barvinsky:2003kg} 
  A.~O.~Barvinsky,
  ``Nonlocal action for long distance modifications of gravity theory,''
  Phys.\ Lett.\ B {\bf 572}, 109 (2003)
  [hep-th/0304229].
  
 
\bibitem{Barvinsky:2011hd} 
  A.~O.~Barvinsky,
  ``Dark energy and dark matter from nonlocal ghost-free gravity theory,''
  Phys.\ Lett.\ B {\bf 710}, 12 (2012)
  [arXiv:1107.1463 [hep-th]].
  
\bibitem{Barvinsky:2011rk} 
  A.~O.~Barvinsky,
  ``Serendipitous discoveries in nonlocal gravity theory,''
  Phys.\ Rev.\ D {\bf 85}, 104018 (2012)
  [arXiv:1112.4340 [hep-th]].
  
\bibitem{Dirian:2014xoa} 
  Y.~Dirian and E.~Mitsou,
  ``Stability analysis and future singularity of the $m^2 R \Box^{-2} R$ model of non-local gravity,''
  JCAP {\bf 1410}, no. 10, 065 (2014)
  [arXiv:1408.5058 [gr-qc]].
  
\bibitem{Nersisyan:2016hjh} 
  H.~Nersisyan, Y.~Akrami, L.~Amendola, T.~S.~Koivisto and J.~Rubio,
  ``Dynamical analysis of $R\dfrac{1}{\Box^{2}}R$ cosmology: Impact of initial conditions and constraints from supernovae,''
  Phys.\ Rev.\ D {\bf 94}, no. 4, 043531 (2016)
  [arXiv:1606.04349 [gr-qc]].
  
\bibitem{Amendola:2017qge} 
  L.~Amendola, N.~Burzilla and H.~Nersisyan,
  ``Quantum Gravity inspired nonlocal gravity model,''               
  Phys.\ Rev.\ D {\bf 96}, no. 8, 084031 (2017)
  [arXiv:1707.04628 [gr-qc]].
  
\bibitem{Foffa:2013vma} 
  S.~Foffa, M.~Maggiore and E.~Mitsou,
  ``Cosmological dynamics and dark energy from nonlocal infrared modifications of gravity,''
  Int.\ J.\ Mod.\ Phys.\ A {\bf 29}, 1450116 (2014)
  [arXiv:1311.3435 [hep-th]].
  
\bibitem{Kehagias:2014sda} 
  A.~Kehagias and M.~Maggiore,
  ``Spherically symmetric static solutions in a nonlocal infrared modification of General Relativity,''
  JHEP {\bf 1408}, 029 (2014)
  [arXiv:1401.8289 [hep-th]].
  
\bibitem{Tsamis:2016boj} 
  N.~C.~Tsamis and R.~P.~Woodard,
  ``Improved cosmological model,''
  Phys.\ Rev.\ D {\bf 94}, no. 4, 043508 (2016)
  [arXiv:1606.06967 [gr-qc]].
  
\bibitem{Tsamis:2005hd} 
  N.~C.~Tsamis and R.~P.~Woodard,
  ``Stochastic quantum gravitational inflation,''
  Nucl.\ Phys.\ B {\bf 724}, 295 (2005)
  [gr-qc/0505115].
  
\bibitem{Tsamis:2009ja} 
  N.~C.~Tsamis and R.~P.~Woodard,
  ``A Phenomenological Model for the Early Universe,''
  Phys.\ Rev.\ D {\bf 80}, 083512 (2009)
  [arXiv:0904.2368 [gr-qc]].
  
\bibitem{Deser:2013uya} 
  S.~Deser and R.~P.~Woodard,
  ``Observational Viability and Stability of Nonlocal Cosmology,''    
  JCAP {\bf 1311}, 036 (2013)
  [arXiv:1307.6639 [astro-ph.CO]].

\bibitem{Kumar:2018chy} 
  U.~Kumar and S.~Panda,
  ``Non-local cosmological models,''
  arXiv:1806.09616 [gr-qc].

\bibitem{Codello:2016xhm} 
  A.~Codello and R.~K.~Jain,
  Int.\ J.\ Mod.\ Phys.\ D {\bf 25}, no. 12, 1644023 (2016)
  doi:10.1142/S0218271816440235
  [arXiv:1605.07630 [gr-qc]].
  
\bibitem{Codello:2016elq} 
  A.~Codello and R.~K.~Jain,
  PoS DSU {\bf 2015}, 008 (2016).
  doi:10.22323/1.268.0008
  
\bibitem{Codello:2016neo} 
  A.~Codello and R.~K.~Jain,
  Eur.\ Phys.\ J.\ C {\bf 78}, no. 5, 357 (2018)
  doi:10.1140/epjc/s10052-018-5839-4
  [arXiv:1603.00028 [gr-qc]].

 
\bibitem{LT:1918}
  Lense, J., and Thirring, H.,
  Phys. Z. 19, 156(1918).

\bibitem{Gravitation}
  Misner, C. W., Thorne, K. S., and Wheeler, J. A.,
  (1973). Gravitation, Freeman, New York,
  sec. 40.7.

\bibitem{Cornell:2017irh} 
  A.~S.~Cornell, G.~Harmsen, G.~Lambiase and A.~Mazumdar,
  ``Rotating metric in nonsingular infinite derivative theories of gravity,''
  Phys.\ Rev.\ D {\bf 97}, no. 10, 104006 (2018)
  [arXiv:1710.02162 [gr-qc]].

\bibitem{nj} 
  E.\ T.\ Newman and A.\ I.\ Janis,
  ``Note on the Kerr Spinning‐Particle Metric,'' 
  J.\ Math.\ Phys. {\bf 6}, 915 (1965).

\bibitem{Demianski:1972uza}
  M.~Demiański,
  ``New Kerr-like space-time,''
  Phys.\ Lett.\ A {\bf 42} (1972) no.2,  157.

\bibitem{Hawking:1973uf}
  S.~W.~Hawking and G.~F.~R.~Ellis,
  ``The Large Scale Structure of Space-Time,''

\bibitem{Everitt:2011hp} 
  C.~W.~F.~Everitt {\it et al.},
  ``Gravity Probe B: Final Results of a Space Experiment to Test General Relativity,''
  Phys.\ Rev.\ Lett.\  {\bf 106}, 221101 (2011)
  [arXiv:1105.3456 [gr-qc]].

\bibitem{Calcagni:2010ab}
  G.~Calcagni and G.~Nardelli,
  ``Non-local gravity and the diffusion equation,''
  Phys.\ Rev.\ D {\bf 82} (2010) 123518
  [arXiv:1004.5144 [hep-th]].

\bibitem{Edholm:2016hbt} 
  J.~Edholm, A.~S.~Koshelev and A.~Mazumdar,
  ``Behavior of the Newtonian potential for ghost-free gravity and singularity-free gravity,''
  Phys.\ Rev.\ D {\bf 94}, no. 10, 104033 (2016)
  [arXiv:1604.01989 [gr-qc]].

\bibitem{Conroy:2014eja} 
  A.~Conroy, T.~Koivisto, A.~Mazumdar and A.~Teimouri,
  ``Generalized quadratic curvature, non-local infrared modifications of gravity and Newtonian potentials,''
  Class.\ Quant.\ Grav.\  {\bf 32}, no. 1, 015024 (2015)
  [arXiv:1406.4998 [hep-th]].
  
\bibitem{Biswas:2013cha} 
  T.~Biswas, A.~Conroy, A.~S.~Koshelev and A.~Mazumdar,
  ``Generalized ghost-free quadratic curvature gravity,''
  Class.\ Quant.\ Grav.\  {\bf 31}, 015022 (2014)
  Erratum: [Class.\ Quant.\ Grav.\  {\bf 31}, 159501 (2014)]
  [arXiv:1308.2319 [hep-th]].
  
\bibitem{Biswas:2013kla} 
  T.~Biswas, T.~Koivisto and A.~Mazumdar,
  ``Nonlocal theories of gravity: the flat space propagator,''
  arXiv:1302.0532 [gr-qc].
  
\bibitem{Erbin:2014aja}
  H.~Erbin,
  ``Deciphering and generalizing Demiański–Janis–Newman algorithm,''
  Gen.\ Rel.\ Grav.\  {\bf 48} (2016) no.5,  56
  [arXiv:1411.2909 [gr-qc]].
  
\bibitem{Erbin:2014aya}
  H.~Erbin,
  ``Janis–Newman algorithm: simplifications and gauge field transformation,''
  Gen.\ Rel.\ Grav.\  {\bf 47} (2015) 19
  [arXiv:1410.2602 [gr-qc]].
  
\bibitem{Buoninfante:2018xif}
  L.~Buoninfante, A.~S.~Cornell, G.~Harmsen, A.~S.~Koshelev, G.~Lambiase, J.~Marto and A.~Mazumdar,
  ``Non-singular rotating metric in ghost-free infinite derivative gravity,''
  arXiv:1807.08896 [gr-qc].

\bibitem{Balasin:1993kf}
  H.~Balasin and H.~Nachbagauer,
  ``Distributional energy momentum tensor of the Kerr-Newman space-time family,''
  Class.\ Quant.\ Grav.\  {\bf 11} (1994) 1453
  [gr-qc/9312028].  

\bibitem{Visser:2007fj}
  M.~Visser,
  ``The Kerr spacetime: A Brief introduction,''
  arXiv:0706.0622 [gr-qc].
  
\bibitem{Adler:1999yt} 
  R.~J.~Adler and A.~S.~Silbergleit,
  ``A General treatment of orbiting gyroscope precession,''
  Int.\ J.\ Theor.\ Phys.\  {\bf 39}, 1291 (2000)
  [gr-qc/9909054].



\end{thebibliography}
\end{document}